\documentclass[11pt,a4paper]{article}
\usepackage{amsmath}
\usepackage{amsthm}
\usepackage{amssymb}
\usepackage{graphicx}
\usepackage[cp1251]{inputenc}
\usepackage[T2A]{fontenc}

\usepackage{graphicx}
\usepackage[all]{xy}

\input amssymb.sty
\newcommand{\beq}[1]{\begin{equation}\label{#1}}
\newcommand{\eq}{\end{equation}}

\newtheorem{defi}{Definition}[section]

\newtheorem{conj}{Conjecture}[section]

\begin{document}
\setcounter{footnote}{1}
\def\thefootnote{\fnsymbol{footnote}}

\begin{flushright}
ITEP-TH-35/15
\end{flushright}

\vspace{1cm}

\begin{center}
{\Large{\sc Phase portraits of the generalized full symmetric Toda systems on rank 2 groups
}}\\

\ \\
Yu.B. Chernyakov\footnote{Institute for Theoretical and Experimental Physics, Bolshaya Cheremushkinskaya, 25,
117218, Moscow, Russia.}$^{,}$\footnote{Joint Institute for Nuclear Research, Bogoliubov Laboratory of Theoretical Physics, 141980, Dubna, Moscow region, Russia.}, chernyakov@itep.ru\\
G.I Sharygin\footnotemark[2]$^{,}$\footnotemark[3]$^{,}$\footnote{Lomonosov
Moscow State University, Faculty of Mechanics and Mathematics, GSP-1, 1 Leninskiye Gory, Main Building, 119991, Moscow, Russia.}, sharygin@itep.ru\\
A.S. Sorin\footnote{Joint Institute for Nuclear Research, Bogoliubov Laboratory of Theoretical Physics and
Veksler and Baldin Laboratory of High Energy Physics, 141980, Dubna, Moscow region, Russia.}$^{,}$\footnote{National Research Nuclear University MEPhI
(Moscow Engineering Physics Institute),
Kashirskoye shosse 31, 115409 Moscow, Russia}
, sorin@theor.jinr.ru
\end{center}

\begin{abstract}
In this paper we continue investigations that we began in our previous works (\cite{CSS14}, \cite{CSS15}), where we proved, that the phase diagram of Toda system on special linear groups can be identified with the Bruhat order on symmetric group, when all the eigenvalues of Lax matrix are distinct, or with the Bruhat order on permutations of a multiset, if there are multiple eigenvalues. We show, that the coincidence of the phase portrait of Toda system and the Hasse diagram of Bruhat order holds in the case of arbitrary simple Lie groups of rank $2$: to this end we need only to check this property for the two remaining groups of second rank, $Sp(4,\mathbb R)$ and the real form of $G_2$.
\end{abstract}

\tableofcontents
\noindent {}
\newpage

\vspace{1.2cm}
\section{Introduction}
As one knows (see \cite{BBR}, \cite{KY}), Toda dynamical system can be generalized to an integrable system on an arbitrary semisimple Lie group; this system is called \textit{generalized (full symmetric) Toda system}. Thus one can ask about the geometric properties (invariant manifolds, singular points, phase portrait etc.) of such system on arbitrary group. In particular, in our papers \cite{CSS14} and \cite{CSS15} we described the phase portrait of this system on the real special linear group (in more abstract terms on the real forms of the $A_n$ series). We showed, that it can be identified with the Hasse diagram of the (strong) Bruhat order on the corresponding Weyl group (or on its generalization, when the eigenvalues are not all distinct). This result generalizes in some sense the classical results of \cite{DNT}; it appeared as an attempt to give strict mathematical interpretation of results of \cite{FS} (see also \cite{FS2} where Toda systems on other Lie groups are considered from similar point of view, and \cite{KW}, in which the geometry of moment map of the full symmetric Toda flow is investigated).

Then the natural question was, whether similar properties still hold for other semisimple Lie groups. The evident conjecture is, that this is true; this is the question, which we attempt to investigate in the present paper. In fact the present article was first conceived as an appendix to the paper \cite{CSS15} (in which we described the phase portrait of the Toda system on the special linear group, in case when the eigenvalues of Lax matrix are not distinct) to support this conjecture. However at some moment we realized that the size of the appendix was nearing that of the paper itself and decided to split the paper into two.

The subject of the present short notice is to give an explicit description of the behavior of Toda flows on remaining real Lie groups of rank $2$, i.e. on $Sp(4,\mathbb R)$ and the real form of $G_2$ (the case of $SL(2,\mathbb R)$ was investigated in details in our first paper, \cite{CSS14}) as further justification of the conjecture we make. Our reasoning is a slight modification of the methods we used in previous papers, however, applying these methods to groups, different from $SL(n,\mathbb R)$ turned out to be somewhat tricky. The reason for this is in the fact, that the definition of Toda system on a Lie group is usually based on the considerations of roots and Chevalley's basis, while our search of phase portrait is based on the large collection of invariant varieties of the flow, which can be described in the terms of flag spaces and matrix representation of the group.

Thus in order to solve the problem we were obliged to look for a suitable reformulation of the definition of the Toda system, so that the matrix representation of the group could be used. It turns out, that such reformulation indeed exists; it was given, in particular in the paper \cite{KY}. In fact, one can show that there is an embedding of a semisimple group $G$ into the special linear group of suitable dimension such that $G$ is preserved by the Toda system on $SL(n,\mathbb R)$; then Toda system on $G$ is equal to the restriction of the system on $SL(n,\mathbb R)$. However, it was not easy to find such an embedding, in particular for the group $G_2$; moreover, after it is chosen one still needs choose correct invariant varieties to be able to describe the phase structure.

We show that in both cases the conjecture, that we make, still holds. Although the result is positive, it is not clear, if the methods we use can be extended further to the groups of higher ranks in $B,\,C,$ and $D$ series, and still less evident, if they can be applied to the groups $E_7,\,E_8$ and $F_4$. Even if this is possible, the computations become more and more complicated as the rank grows, so one should look for more powerful tools.

The composition of this paper is as follows. In the first few sections we briefly recall basic facts that we are going to use; we begin with the root systems, Weyl groups, and Bruhat order; for more details the reader should refer to the books \cite{Fulton}, \cite{Bjorn}, \cite{Br} (see also our papers \cite{CSS14,CSS15}). We further give a brief excursion into the basic properties of Toda system. We describe it both in terms of the Chevalley basis and as the restriction of the full symmetric system from $SL(n,\mathbb R)$ to the embedded group (under proper conditions onto this embedding) and recall the definition of \textit{minor varieties}, which are invariant sets of Toda system. We also recall that Toda system can be regarded as a gradient flow on a suitable flag space and remind basic properties of gradient flows on manifolds (one can call it \textit{elementary Morse theory}). In remaining two sections we apply the methods to the groups $Sp(4,\mathbb R)$ and $G_2$ respectively; to this end we show how the embedding of the necessary type is defined for these groups, and in what way the minor varieties should be taken so that the phase portrait can be drawn.

\section{Preliminary facts}
In this section we collect the facts necessary for further work. We try to make exposition as brief and as self-contained as possible (although these aims are somewhat contradictory). So we restrict our attention to most important definitions and facts, totally omitting the proofs and definitions that we shall not use in our paper; for example we ignore Dynkin diagrams and their r\^ole in the classification of Lie algebras. For further information on this and other subjects the reader may refer to the papers and books that are in our list of references.
\subsection{Root systems and Weyl groups}
\label{sec:roots}
An account of root systems in the context of Lie groups theory and its relation to Weyl groups can be found in many books; for example one can found it in \cite{Bjorn, Br}. Here we give bare basics of it.

As one knows, $1$-connected simple complex Lie groups can be classified in terms of their root systems. Namely, choose a maximal commutative subalgebra $\mathfrak h\subseteq\mathfrak g$ of the Lie algebra (Cartan subalgebra) of $\mathfrak g$; $\mathrm{dim}\,\mathfrak h=r$ is called \textit{the rank of $\mathfrak g$}. Then there exists a collection of linear functionals $\alpha_1,\dots,\alpha_{n-r}$ on $\mathfrak h$ such that a basis $\mathfrak g$ can be chosen, which will consist of a basis $H_1,\dots,H_r$ of $\mathfrak h$ and a collection of vectors $E_{\alpha_1},\dots,E_{\alpha_{n-r}}\in\mathfrak g\setminus\mathfrak h$ so that the following commutation relations hold
\[
[H_i,H_j]=0,\, [E_\alpha,H_i]=\alpha(H_i)E_\alpha,\ [E_\alpha,E_\beta]=n_{\alpha\beta}E_{\alpha+\beta},\ \mbox{if}\ \alpha+\beta\neq0,\ [E_\alpha,E_{-\alpha}]\in\mathfrak h.
\]
Here $\alpha,\beta$ are some functionals from the list we chose above, $n_{\alpha\beta}\in\mathbb Z$ and we assume in the third formula that $\alpha+\beta$ is again in the list, and $\alpha([E_\alpha,E_{-\alpha}])\ne0$. Basis with this property is called \textit{Cartan-Weyl basis}.

The functionals $\alpha_i$ from this set are called \textit{roots} of the Lie algebra; they are uniquely defined by the conditions above (and by the choice of Cartan algebra). As one can see, they are not in general linearly independent. If a basis in the functional space $\mathfrak h^*$ is chosen, one can introduce an ordering on the set of roots as well as divide the roots into positive and negative subsets. One says, that a (positive) root $\alpha$ is \textit{simple}, if $\alpha$ is not equal to a sum of two other roots from the same system. It turns out, that there always exist exactly $r$ simple roots $\alpha_1,\dots,\alpha_r$ in any root system, and the corresponding basis $H_{\alpha_1},\dots,H_{\alpha_r}$ of $\mathfrak h$ given by the condition
\[
\alpha_j(H)=B(H_{\alpha_j},H),\ H\in\mathfrak h,
\]
where $B(X,Y)$ is the Killing form on $\mathfrak h$, so that the relations above are the same, except that:
\[
[E_{\alpha_i},E_{-\alpha_i}]=H_{\alpha_i}
\]
for all simple roots $\alpha_1,\dots,\alpha_r$. The basis with these properties is called \textit{Chevalley basis} of the Lie algebra $\mathfrak g$.

Another important notion, closely related to the Cartan subalgebra of a Lie algebra is \textit{Weyl group}. This group has many definitions, including purely combinatoric ones. One of the convenient ways to define it is as follows: consider the adjoint action of the Lie group $G$ on its Lie algebra $\mathfrak g$. For a given cartoon subalgebra $\mathfrak h\subseteq\mathfrak g$ we define its \textit{normalizer} $N_G(\mathfrak h)\subseteq G$ as the stabilizer of $\mathfrak h$ under this action. Clearly, this subgroup $N_G(\mathfrak h)$ contains the commutative subgroup $T_\mathfrak h$ of $G$, spanned by $\mathfrak h$, as its normal subgroup. Then
\[
W_G=N_G(\mathfrak h)/T_\mathfrak h;
\]
in fact, one can define $W_G$ in this way for any maximal commutative subgroup $T_\mathfrak h$ instead of $\mathfrak h$. It turns out that $W_G$ of a semisimple Lie group is discrete and finite. It follows from the definition that $W_G$ comes equipped with a natural action on $\mathfrak h$; this action allows one identify $W_G$ with a group of linear transformations of $\mathfrak h$, induced by symmetries with respect to the orthogonal complements of vectors $H_{\alpha_i}$. One can define the Weyl groups in this way abstractly as the special kind of Coxeter reflection groups.

\subsection{Bruhat order and Bruhat cell decomposition}
\label{sec:Bru}
The main references for this section, as well as for the previous one are the books \cite{Br} and \cite{Bjorn}; see also \cite{Fulton}.

Any Coxeter reflection group $W$ has a canonical system of generators $t_i$, corresponding to the ``basic'' reflections, which generate $W$. It follows that one can use this system to define various additional structures on $W$. First of all, one can define \textit{length $l(g)$} of an element $g\in W$
\[
l(g)=\min\{n\mid g=t_{i_1}\dots t_{i_n}\}.
\]
This function can be used to define a partial order on $W$. A more subtle partial order is the so called \textit{strong Bruhat order}, which can be defined as the closure (in the sense of partial orders) of the following elementary relation:
\[
u\prec v\Leftrightarrow v=tu,\ \mbox{for some elementary reflection}\ t.
\]
A convenient way to represent this order is by drawing a graph, which is called the Hasse diagram of the order, or simply \textit{Bruhat graph}.  This is a directed graph, whose vertices correspond to the elements of the group and edges connect those points $u$ and $v$, for which the elementary relation $u\prec v$ holds; the corresponding edge is then directed from $u$ to $v$ (more accurately, Hasse diagram should also contain the edges, that correspond to the consequences of elementary relations, which we shall usually omit).

Since the groups $W_G$ are Coxeter groups, we see, that in particular Bruhat order can be introduced on all Weyl groups. In some cases this order can be given a clear geometric meaning. For instance, in the case of the groups $SL(n,\mathbb C)$ ($A$-series of the simple groups), Weyl groups are equal to the full permutation groups $S_n$. It turns out, that there is a geometric way to introduce Bruhat order here. Namely, one can consider the \textit{(real) flag space}
\[
F_n(\mathbb R)=\{\{0\}=V_0\subset V_1\subset\dots\subset V_n=\mathbb R^n\mid \mathrm{dim}\,V_i=i\}=SL(n,\mathbb R)/B^+_n,
\]
where $B^+_n$ is the group of upper triangular matrices with unit determinant. One can introduce a cell decomposition of this space into $n!$ cells $X_v$, called \textit{Schubert cells}, enumerated by permutations, elements $v\in S_n$; this cell decomposition is called \textit{Schubert cell decomposition} or \textit{Bruhat cell decomposition}.

Schubert cells are open varieties in $F_n(\mathbb R)$, determined by conditions on the ranks of intersections of the corresponding spaces $V_i$ with the given (coordinate) hyperplanes in $\mathbb R^n$. One can show, that in this case $v\prec u$ in the sense of Bruhat order on $S_n$ if and only if $X_v\subset \bar X_u$, where $\bar X_u$ is the closure of the cell $X_u$.

One can look for similar interpretations of Bruhat order for other classical and sporadic groups. It turns out, that there is an analogue of flag spaces:
\[
F(G)=G/B^+,
\]
where $B^+$ is a positive Borel subgroup of $G$. There is also an analog of Bruhat cell decomposition of $F(G)$ given by the orbits of $B^+$, enumerated by the elements in $W_G$. But its relation with Bruhat order on $W_G$ is not that obvious, especially in the case of sporadic Lie groups.

\subsection{Toda system}
\label{sec:genTod}
The literature about Toda flow is quite abundant. In our treatment here we follow \cite{KY} and \cite{deMPe}. Other details, in particular the integrability etc. can be found in \cite{DLNT} and \cite{EFS}; the Adler-Kostant-Symes scheme was developed in \cite{Ad,K1} and \cite{S}.

In the theory of Toda flow an important r\^ole is played by the Toda systems on classical simple groups, or rather on the  corresponding Lie algebras (in fact, one can speak about the induced system on the corresponding flag spaces $G/B^+$, like in the usual case). There are many ways, in which the system on these groups can be defined. For example, one can give an explicit formula for the $L$-matrix and $M$-matrix in terms of the canonical root system (see \cite{KY}): for any Cartan-Wayl basis $H_1,\dots,H_r,\,E_{\alpha_1},\dots,E_{\alpha_{n-r}}$, where $r=\mathrm{rk}\,\mathfrak g,\ n=\mathrm{dim}\,\mathfrak g$ let $\Delta^+$ denote the subset of positive roots (with respect to a given basis), then
\[
\begin{aligned}
L&=\sum_{i=1}^ra_iH_i+\sum_{\alpha\in\Delta^+}b_\alpha(E_\alpha+E_{-\alpha}),\\
M&=\sum_{\alpha\in\Delta^+}b_\alpha(E_\alpha-E_{-\alpha});
\end{aligned}
\]
and we obtain the following differential equation:
\begin{equation}
\label{eq1}
\dot{L}=[M,L].
\end{equation}
Another possible approach is to consider the Iwasawa decomposition of the group and apply Adler-Kostant-Symes scheme to it. In both case, when $G=SL(n,\mathbb R)$, we obtain the following matrix representations:
\beq{LM}
L =
\begin{pmatrix}
 b_{11} & b_{12} & b_{13} & ... & b_{1n}\\
 b_{12} & b_{22} & b_{23} &... & b_{2n}\\
 ... & ... & ... & ...& ...\\
 b_{1n} & b_{2n} & b_{3n} &  ... & b_{nn}
\end{pmatrix},\ \
M=\begin{pmatrix}
 0 & b_{12} & b_{13} & ... & b_{1n}\\
 -b_{12} & 0 & b_{23} &... & b_{2n}\\
 ... & ... & ... & ...& ...\\
 -b_{1n} & -b_{2n} & -b_{3n} &  ... & 0
 \end{pmatrix}=L_+-L_-,
\eq
where $L_+$ and $L_-$ are the upper and lower diagonal parts of $L$ respectively.

In the present paper we choose the point of view on the generalized Toda system on a Lie group that was used in the paper \cite{FS2} (see also \cite{KY}). It is based on the following idea, generalizing the Adler-Kostant-Symes scheme: \textit{if we embed the group $G$ into a suitable $SL(n,\,\mathbb R)$ so that the Cartan subalgebra is mapped into the diagonal matrices and root vectors correspond to the upper or lower triangular matrices, so that the matrices, corresponding to the $+\alpha$ and $-\alpha$ vectors are transpose of each other; then the Toda equation is given by the formula \eqref{LM}, where Lax matrix $L$ is from the intersection of the symmetric matrices and the image of $\mathfrak{g}\subset\mathfrak{sl}_n$, snd $M$ is given by the restriction of the anti-symmetrization procedure to $L$.} In effect, one can even choose embedding in which positive root vectors will be represented by strictly upper-triangular matrices, but this is more, than we need for our purposes.

One can show, that the system, obtained in this way is a Hamiltonian integrable system. Another important property of the Toda system is that in some sense it also has the structure of a gradient flow. There are many ways to describe it (see for instance \cite{deMPe}, \cite{BBR}). Here we present one, adapted for the construction of Toda flow we use.

Namely, the embedding (the real form of) $G$ into a special linear group $SL(n,\mathbb R)$ as explained above, then the maximal compact subgroup of $G$ will be mapped inside the group $SO(n,\mathbb R)$ of orthogonal matrices. Besides this, it will follow from equation\eqref{eq1} that the eigenvalues of the Lax matrix are preserved by Toda flow. Since every real symmetric matrix $L$ can be represented in the form $\Psi\Lambda\Psi^t$, where $\Psi$ is orthogonal and $\Lambda$ is diagonal eigenvalues matrix, we can use the same decomposition for the Lax matrix $L$ of the Toda system; in this case $\Psi$ will be a matrix from the maximal compact subgroup $K_G\subset G$, and $\Lambda$ will be a matrix from the Cartan subgroup of $G$.

Thus, since the matrix $\Lambda$ does not change under the action of Toda flow, fixing it we obtain a dynamical system for $\Psi$
\begin{equation}
\label{eq2}
\frac{d\Psi}{dt}=M(\Psi)\Psi,\ M(\Psi)=(\Psi\Lambda\Psi^t)_+-(\Psi\Lambda\Psi^t)_-.
\end{equation}
This system \eqref{eq2} is in certain sense equivalent to the Toda flow (in fact, Toda flow is obtained from adjoint action of $\Psi(t)$ on $\Lambda$). In fact, one can consider this system on a flag space, associated with $G$, which is equal to
\[
F(G)=K_G/(K_G\bigcap H),
\]
where $H$ is the chosen maximal commutative subgroup.

Now one can show that equations \eqref{eq2} are in fact given by a gradient flow on $K_G$ (or on $F(G)$). To this end one introduces an invariant Euclidean structure on $\mathfrak{so}_n$ (a suitable deformation of the Killing form), and extends it to a Riemmanian on $SO(n,\mathbb R)$; this structure is then restricted to $K_G$, embedded into the orthogonal group as explained above. Then one can show, that the equation \eqref{eq2} has the form of the gradient flow with respect to the chosen Riemannian structure of the following function:
\[
F_G(\Psi)=Tr(\Psi\Lambda\Psi^tN),
\]
where $\Lambda$ is the matrix of eigenvalues, and $N$ a suitable diagonal matrix (representing an element in the chosen Cartan subgroup). List of such elements for different groups can be found for instance in paper \cite{BBR}.

Another important property of the system \eqref{eq2} is that one can find many varieties in $SO(n,\mathbb R)$, preserved by this system. An important large collection of such invariant sets is given by so-called \textit{minor surfaces} (see \cite{CS} for details): they are the null-sets of the functions $M^+_I(g),\ M^-_J$ (where $I,\,J$ are poly-index sets $I=(1\le i_1<i_2<\dots<i_p\le n),\ J=(1\le j_1<j_2<\dots<j_q\le n)$ and $g\in SO(n,\mathbb R)$), given by the formulas
\[
M^+_I(g)=\det(g^I_{1,2,\dots,p}),\ M^-_J(g)=\det(g^J_{n-q+1,\dots,n-1,n}),
\]
where $g^I_{1,2,\dots,p})$ and $g^J_{n-q+1,\dots,n-1,n}$ are the sub matrices of $g$, spanned by the first $p$ (resp. last $q$) rows and the columns, given by $I$ (resp by $J$). Below we shall make a wide use of these minor surfaces; based on the fact, that intersection of these surfaces with embedded group $G$ are preserved by equation \eqref{eq2}: this follows from the fact, that minor surfaces are preserved by all equations of this form.

Another important family of invariant submanifolds of Toda flow on $SL(n,\mathbb R)$ (or rather on the corresponding flag space) consists of Schubert varieties (see section \ref{sec:Bru}). These varieties are also invariant with respect to arbitrary flow of the form \eqref{eq2}. However, one should be cautious in this respect: equation \eqref{eq2} can be regarded both as a dynamical system on the group $G$, embedded into $SL(n,\mathbb R)$ and as a dynamical system on $SL(n,\mathbb R)$ itself. It is as the latter it has Schubert cells as invariant submanifolds. Of course, intersections of these sub manifolds with $G$ are invariant subsets of the Toda system on $G$; however, it is not clear, what is the relation of these intersections and Bruhat cells of $G$ itself.

\subsection{Elementary Morse theory}
The notion of Morse function and its properties can be found in numerous books. Its application to Toda system was considered by us in \cite{CSS14} and \cite{CSS15}.

The property of being gradient flow makes the problem of describing the phase portrait of Toda system much easier: as one knows, gradient flows cannot have closed trajectories; trajectories gradient flows on compact manifoldds (in particular on flag space $F(G)$) always connect singular points of the flow (i.e. the points, in which the vector field vanishes). Besides this, the values of the function $f$ always grow in the direction of the gradient; in particular, if a function $f$ has only two singular points with different values on a manifold, all the trajectories of the flow go from the point in which the value of $f$ is less to the point, in which it is greater.

A particularly good type of the gradient flows are the flows of Morse functions. Let us recall the definitions:
\begin{defi}
Let $f$ be a smooth real-valued function on a manifold $M$; a point $x_0\in M$ is called singular for $f$ if the differential of $f$ vanishes at $x_0$. A singular point $x_0$ of $f$ is called non-degenerate if the matrix $\left(\frac{\partial^2 f}{\partial x_i\partial x_j}\right)$ at $x_0$ is non-degenerate, i.e. if its determinant (Hessian of $f$ at $x_0$) does not vanish. Finally, the function $f$ on a smooth compact manifold $M$ is called Morse function, if it has only a finite number of nondegenerate singular points on $M$ (sometimes an additional condition, that the values of $f$ at these points are distinct).
\end{defi}
Morse functions on smooth manifolds, their structure and relation to the topology of manifolds are studied by Morse theory. We shall need only few facts from this theory.

As any nondegenerate real quadratic form, the matrix $\left(\frac{\partial^2 f}{\partial x_i\partial x_j}\right)$ at $x_0$ is completely (up to a linear change of coordinates) characterized by the number of positive and negative terms in its canonic form. Their difference is called \textit{index of the singular point $x_0$}. One can prove, that there always exist local coordinates $(\xi_1,\dots,\xi_n)$ on $M$ centered at $x_0$, such that the function $f$ near $x_0$ has the following form enthuse coordinates:
\[
f(\xi_1,\dots,\xi_n)=f(x_0)+\xi_1^2+\dots+\xi_k^2-\xi_{k+1}^2-\dots-\xi_n^2,
\]
where $2k-n$ is the index of $x_0$. Moreover, these coordinates can be made orthogonal at $x_0$ with respect to a given Riemannian structure (though not necessarily orthonormal).

The subspace of $T_{x_0}M$, spanned by the basis vectors $\partial_{\xi_1},\dots,\partial_{\xi_k}$ is the tangent space of the local submanifold $W^+_{x_0}\subseteq M$, spanned by the trajectories of the gradient flow of $f$, leaving $x_0$; this submanifold is called \textit{stable variety} of $x_0$. Similarly, the subspace of $T_{x_0}M$, spanned by $\partial_{\xi_{k+1}},\dots,\partial_{\xi_n}$ is target to the submanifold $W^-_{x_0}$, spanned by the trajectories, abutting at $x_0$, which is called \textit{unstable variety} of $x_0$. This information can often be crucial in determining the phase structure of a flow: if we know for instance that there there are $2$ two negative directions in a canonical coordinate system at $x_0$ and that, on the other hand, $x_0$ is one of two singular points of the flow in an invariant $2$-dimensional submanifold $\Sigma\subset M$, then we can conclude that all trajectories, leaving $x_0$ abut at the second point in $\Sigma$.

\section{Two examples}
This section deals with the geometric properties of Toda flow on groups $Sp(4,\mathbb R)$ and (the real form of) the simplest exceptional group $G_2$. In our research here we freely (and often without references) use the facts listed in the previous section. Similar computations can be found, for instance in \cite{FS,FS2}.
\subsection{The group $Sp(4,\mathbb R)$}
Recall, that $Sp(2n,\mathbb R)$ is the group of orientation-preserving linear transformations of $\mathbb R^{2n}$, preserving a given nondegenerate antisymmetric bilinear form $J$, i.e.
\[
Sp(2n,\mathbb R)=\{A\in SL(2n,\mathbb R)\mid AJ=JA\}.
\]
One usually takes $J=\begin{pmatrix}0 & I_n\\ -I_n & 0\end{pmatrix}$, where $I_n$ is the $n\times n$ unit matrix. The flag space of this group can be described in the terms of symplectic subspaces in $\mathbb R^{2n}$.

Let us now restrict our attention to $Sp(4,\mathbb R)$. As one knows, all antisymmetric nondegenerate forms are equivalent up to the choice of basis. But if we want to embed $Sp(4,\mathbb R)$ into $SL(4,\mathbb R)$ so as to make positive roots go into upper triangular matrices, we should take
\[
J=\begin{pmatrix}
0 & 0 & 0 & 1\\
0 & 0 & 1 & 0\\
0 &-1 & 0 & 0\\
-1& 0 & 0 & 0
\end{pmatrix}.
\]
From now on we shall mutely assume that such representation is taken, when we speak about symplectic matrices. For instance, under this embedding Cartan algebra of $\mathfrak{sp}(4,\mathbb R)$ is diagonal and its Chevalley basis is given by
\beq{tN40}
\begin{array}{c}
h_{1} = \left(
\begin{array}{cccc}
 1 & 0 & 0 & 0 \\
 0 & 0 & 0 & 0 \\
 0 & 0 & 0 & 0 \\
 0 & 0 & 0 & -1
\end{array}
\right), \ \ \
h_{2} = \left(
\begin{array}{cccc}
 0 & 0 & 0 & 0 \\
 0 & 1 & 0 & 0 \\
 0 & 0 & -1 & 0 \\
 0 & 0 & 0 & 0
\end{array}
\right),
\end{array}
\eq
and the Lie algebra of the maximal compact subgroup in $Sp(4,\mathbb R),\ U(2)$ is spanned by the matrices of the form
\beq{tN4}
\Theta = \left(
\begin{array}{cccc}
 0 & \theta _1 & \theta _3 & -\theta_4 \\
 -\theta _1 & 0 & \theta _2 & \theta_3 \\
 -\theta _3 & -\theta _2 & 0 & -\theta_1 \\
 \theta _4 & -\theta _3 & \theta _1 & 0
\end{array}
\right).\\
\eq
It may also be instructive to take a closer look at the coordinate form of the Toda flow equations under this embedding. The Lax matrix here is a symmetric matrix inside the $\mathfrak{sp}(4,\mathbb R)$:
\begin{align}
\label{aN4}
\qquad&\qquad L(U)=U \Lambda U^{-1}=\begin{pmatrix}
 b_{1} & a_{1} & a_{2} & -a_{4}\\
 a_{1} & b_{2} & a_{3} & a_{2}\\
 a_{2} & a_{3} & -b_{2} & -a_{1}\\
 -a_{4} & a_{2} & -a_{1} & -b_{1}
\end{pmatrix}.\\
\intertext{Here $\Lambda$ is the diagonal matrix of eigenvalues of $L$, and $U$ is a matrix from $U(2)=Sp(4,\mathbb R)\bigcap SO(4,\mathbb R)$, the maximal compact subgroup of $Sp(4,\mathbb R)$. Then the $M$ matrix from equation \eqref{LM} is equal to}
%
\label{aN5}
M(U)&=(U \Lambda U^{-1})_{>0} - (U \Lambda U^{-1})_{<0}=\begin{pmatrix}
 0 & a_{1} & a_{2} & -a_{4}\\
 -a_{1} & 0 & a_{3} & a_{2}\\
 -a_{2} & -a_{3} & 0 & -a_{1}\\
 a_{4} & -a_{2} & a_{1} & 0
 \end{pmatrix}.
\end{align}
%
Then the equation, verified by $U$ is
\begin{equation}
\label{N4-1}
\begin{aligned}
&\qquad\qquad\begin{pmatrix}
 u_{11} & u_{12} & u_{13} & u_{14}\\
 u_{21} & u_{22} & u_{23} & u_{24}\\
 u_{31} & u_{32} & u_{33} & u_{34}\\
 u_{41} & u_{42} & u_{43} & u_{44}
\end{pmatrix}'=\\
& = 
\begin{pmatrix}
   (-b_{1} + \lambda_{1})u_{11} & (-b_{1} + \lambda_{2})u_{12} & (-b_{1} - \lambda_{2})u_{13} & (-b_{1} - \lambda_{1})u_{14}\\
       &  &  & \\
(-b_{2} + \lambda_{1})u_{21} - & (-b_{2} + \lambda_{2})u_{22} - & (-b_{2} - \lambda_{2})u_{23} - & (-b_{2} - \lambda_{2})u_{24} -\\
       - 2a_{1}u_{11} & - 2a_{1}u_{12} & - 2a_{1}u_{13}  & - 2a_{1}u_{14}\\
       &  &  & \\
(+b_{2} + \lambda_{1})u_{31} - & (+b_{2} + \lambda_{2})u_{32} - & (+b_{2} - \lambda_{2})u_{33} - & (+b_{2} - \lambda_{1})u_{34} -\\
- 2a_{2}u_{11} - 2a_{3}u_{21} & - 2a_{2}u_{12} - 2a_{3}u_{22} & - 2a_{2}u_{13} - 2a_{3}u_{23} & - 2a_{2}u_{14} - 2a_{3}u_{24}\\
       &  &  & \\
   (-b_{1} - \lambda_{1})u_{41} & (-b_{1} - \lambda_{2})u_{42} & (-b_{1} + \lambda_{2})u_{43} & (-b_{1} + \lambda_{1})u_{44}
\end{pmatrix}\\
%
%
 & =
 \begin{pmatrix}
   (-b_{1} + \lambda_{1})u_{11} & (-b_{1} + \lambda_{2})u_{12} & (-b_{1} - \lambda_{2})u_{13} & (-b_{1} - \lambda_{1})u_{14}\\
       &  &  & \\
(b_{2} - \lambda_{1})u_{21} + & (b_{2} - \lambda_{2})u_{22} + & (b_{2} + \lambda_{2})u_{23} + & (b_{2} + \lambda_{1})u_{24} +\\
 + 2a_{3}u_{31} + 2a_{2}u_{41} &  + 2a_{3}u_{32} + 2a_{2}u_{42} & + 2a_{3}u_{33} + 2a_{2}u_{43} & + 2a_{3}u_{34} + 2a_{2}u_{44}\\
       &  &  & \\
(-b_{2} - \lambda_{1})u_{31} + & (-b_{2} - \lambda_{2})u_{32} + & (-b_{2} + \lambda_{2})u_{33} + & (-b_{2} + \lambda_{1})u_{34} +\\
+ 2a_{1}u_{41} & + 2a_{1}u_{42} & + 2a_{1}u_{43}  & + 2a_{1}u_{44}\\
       &  &  & \\
   (-b_{1} - \lambda_{1})u_{41} & (-b_{1} - \lambda_{2})u_{42} & (-b_{1} + \lambda_{2})u_{43} & (-b_{1} + \lambda_{1})u_{44}\\
\end{pmatrix}
\end{aligned}
\end{equation}
Let us fix the eigenvalues of $L$ to be $0<\lambda_1<\lambda_2$, so that the corresponding element in Cartan algebra is $\lambda_1h_1+\lambda_2h_2$. Since Toda system on $Sp(4,\mathbb R)$ is given by the restriction of equation \eqref{LM} from $SL(4,\mathbb R)$, one can describe the stable points of the flow on the flag space; they are given by the equivalence classes of orthogonal matrices in $Sp(4,\mathbb R)$, i.e. from the intersection $SO(4,\mathbb R)\bigcap Sp(4,\mathbb R)=U(2)$, which preserve the Cartan subalgebra of $Sp(4,\mathbb R)$, i.e. the diagonal matrices from $\mathfrak{sp}(4,\mathbb R)$ are sent into diagonal matrices again. There are eight permutation matrices, that fall into $U(2)$, we shall denote them by $\tilde s_{\mu_1,\mu_2,\mu_3,\mu_4}$, where each $\mu_i=\lambda_1$ or $\lambda_2$ (i.e. we we use the notation of multiset permutation, see \cite{CSS15}). We shall also use the  simple notation $\tilde s_i,\,i=1,\dots,8$. The list of matrices $\tilde{s_i}$ can be found in \cite{FS} pp.34-35.

We can now use the formulas from \cite{KY}, which express the Morse function of Toda system on symplectic flags in terms of the roots of $Sp(n,\mathbb R)$. The Morse function $F_n$ on $Sp(n,\mathbb R)$ is equal to $Tr(LN)$, where the matrix $N$ is determined by the following equations
\begin{align}
N &= (-1)^{1/2}\sum_{j}x_{j}h_{j}\\
\intertext{where we have $n=2l$ and}
x_{i}&=-1/2 \cdot i \cdot (2l-i).\\
\intertext{In particular, when $l=2$ (i.e. for $Sp(4,\mathbb R)$) there are only $2$ elements $h_1,\,h_2$ and in this case we shall have}
x_{1}&=-3/2,\ x_{2}=-2.
\end{align}
This gives, up to a constant multiple:
$$
N=\begin{pmatrix} 0 & 0 & 0& 0\\ 0 & -1 & 0 & 0\\ 0 & 0 & 7 & 0\\ 0 & 0 & 0 & 6\end{pmatrix}.
$$
We are going to express the quadratic part of $F_{Sp(4,\mathbb R)}$ in the singular points $\tilde{s_i},\ i=1,\dots,8$ in terms of the local coordinates, transferred from the Lie algebra (we move it to the corresponding point by left translations). To this end we replace matrix $U$ in a neighborhood of $\tilde{s_i}$ by
\beq{decomppsi}
U=\tilde s_i\exp(\Theta)=\tilde{s_i}+\Theta\tilde{s_i}+\frac12\Theta^2\tilde{s_i}+o(\Theta^2).
\eq
Here $o(\Theta^2)$ denotes the sum of terms of degree 3 and higher in $\theta_i$. Thus the Lax matrix in the neighborhood of the stable points is given by the formula
\beq{decomplax}
L(U)=U \Lambda U^{-1}=\tilde{s_i}\Lambda\tilde{s_i}^{-1}+ [\Theta, \tilde{s_i}\Lambda\tilde{s_i}^{-1}] - \Theta \tilde{s_i}\Lambda\tilde{s_i}^{-1} \Theta + \frac{1}{2}[\Theta^{2}, \tilde{s_i}\Lambda\tilde{s_i}^{-1}]_{+}+o(\theta^2).
\eq
Following is the example of quadratic part of $F_{Sp(4,\mathbb R)}$ at the singular points:
\beq{Hess}
\begin{array}{c}
d^{2}_{\tilde s_{1,2,-2,-1}}F_{Sp(4,\mathbb R)}=\theta _1^2 \left(\lambda _2-\lambda _1\right)+8 \theta _2^2 \lambda _2+7 \theta _3^2
   \left(\lambda _1+\lambda _2\right)+6 \theta _4^2 \lambda _1,\\
\ \\
\lambda _2 > \lambda _1 > 0.
\end{array}
\eq
As one can see, the function $F_{Sp(4,\mathbb R)}$ is indeed a Morse function (moreover the Hessian is diagonal when one chooses the coordinates $\theta_i$), so we can apply the Morse theory again (just as we did it in the paper \cite{CSS14}).However, this information is not sufficient to restore the precise picture of the trajectories connecting the singular points. So, to glean more information, we remark, that since equation \eqref{N4-1} is restriction of the equation \eqref{LM} to the embedded group, its invariant varieties are equal to the intersections of the subgroup with minor surfaces. The following table lists the singular points of our system, their Morse indices (including the signs, standing at the squares of the coordinates $\theta_i$ in the corresponding Hessians) and the list of the minor surfaces, to which they belong:
\begin{equation*}
\footnotesize{
\begin{tabular}{|c|c|c|c|c|}
\hline\cline{1-0}
$$ & $$ & $$ & $$ & $$\\
$i$ & $\tilde{s_i}$ & $\theta _1,\theta _2,\theta _3,\theta _4$ & Index & Minors, $u_{ij}$\\

$$ & $$ & $$ & $$ & $$\\
\hline\cline{1-0}
$$ & $$ & $$ & $$ & $$\\
$1$ & $\tilde s_{1,2,-2,-1}$ & $+,+,+,+$ & $0$ & $u_{12},u_{13},u_{14},u_{41},u_{42},u_{43}$\\

$$ & $$ & $$ & $$ & $$\\
\hline\cline{1-0}
$$ & $$ & $$ & $$ & $$\\
$2$ & $\tilde s_{-1,-2,2,1}$ & $-,-,-,-$ & $4$ & $u_{11},u_{12},u_{13},u_{42},u_{43},u_{44}$\\

$$ & $$ & $$ & $$ & $$\\
\hline\cline{1-0}
$$ & $$ & $$ & $$ & $$\\
$3$ & $\tilde s_{-1,2,-2,1}$ & $+,+,+,-$ & $1$ & $u_{11},u_{12},u_{13},u_{42},u_{43},u_{44}$\\

$$ & $$ & $$ & $$ & $$\\
\hline\cline{1-0}
$$ & $$ & $$ & $$ & $$\\
$4$ & $\tilde s_{1,-2,2,-1}$ & $-,-,-,+$ & $3$ & $u_{12},u_{13},u_{14},u_{41},u_{42},u_{43}$\\

$$ & $$ & $$ & $$ & $$\\
\hline\cline{1-0}
$$ & $$ & $$ & $$ & $$\\
$5$ & $\tilde s_{2,1,-1,-2}$ & $-,+,+,+$ & $1$ & $u_{11},u_{13},u_{14},u_{41},u_{42},u_{44}$\\

$$ & $$ & $$ & $$ & $$\\
\hline\cline{1-0}
$$ & $$ & $$ & $$ & $$\\
$6$ & $\tilde s_{2,-1,1,-2}$ & $-,-,+,+$ & $2$ & $u_{11},u_{13},u_{14},u_{41},u_{42},u_{44}$\\

$$ & $$ & $$ & $$ & $$\\
\hline\cline{1-0}
$$ & $$ & $$ & $$ & $$\\
$7$ & $\tilde s_{-2,1,-1,2}$ & $+,+,-,-$ & $2$ & $u_{11},u_{12},u_{14},u_{41},u_{43},u_{44}$\\

$$ & $$ & $$ & $$ & $$\\
\hline\cline{1-0}
$$ & $$ & $$ & $$ & $$\\
$8$ & $\tilde s_{-2,-1,1,2}$ & $+,-,-,-$ & $3$ & $u_{11},u_{12},u_{14},u_{41},u_{43},u_{44}$\\

$$ & $$ & $$ & $$ & $$\\
\hline
\end{tabular}
\
}
\end{equation*}
By comparing the Morse indices of the points and the sets of points inside various invariant subvarieties (minor surfaces), we can reconstruct the phase picture of this system. Thus we obtain the following picture:
\[
\xymatrix{
                                       & \bullet \\
\bullet\ar@{->}[ur]                  & &\bullet\ar@{->}[ul]\\
\bullet\ar@{->}[u]\ar@{->}[urr] & &\bullet\ar@{->}[u]\ar@{->}[ull]\\
\bullet\ar@{->}[u]\ar@{->}[urr] & &\bullet\ar@{->}[u]\ar@{->}[ull]\\
                           &\bullet\ar@{->}[ul]\ar@{->}[ur]
}
\]
This is the diagram of the $1$-parameter families of trajectories that connect the singular points which correspond $\tilde{s_i}$. As one sees, it coincides with the diagram of Bruhat order for the Weyl group of $Sp(4,\mathbb R)$, found at figure 8.3 on p.250 of \cite{Bjorn}. Also remark, that indices of the singular points coincide with lengths of the corresponding Weyl elements (in the sense of the definitions, given in section \ref{sec:Bru}); this was also the case in all the previous situations we considered in papers \cite{CSS14,CSS15}.

\subsection{The group $G_2$}
The smallest sporadic group from the classification list of simple Lie groups is the group $G_2$ (we shall always consider the real form of all classical groups, thus here we deal with what is usually described as $G_2(\mathbb R)$ rather than with the corresponding complex group, that usually appears in the list). There are many ways to introduce it. For instance, it is the subgroup of $SL(7,\mathbb R)$, which preserves a given symmetric $2$-form and a cubic form on $\mathbb R^7$; alternatively, one can describe it as the group of orthogonal transformations of the $7$-dimensional Euclidean space, which preserves a kind of vector-product on this space (that is is induced from the structure of Caylley numbers).

In this paper we shall base our considerations on the description of $G_2$, given by Gross in \cite{Gro}. It is our purpose to embed this group into $SL(7,\mathbb R)$, so that all the conditions listed in section \ref{sec:genTod} hold. To this end we first construct the $7$-dimensional representation of the corresponding Lie algebra, that will verify similar conditions. Recall, that $\mathfrak g_2$ is described in terms of roots as a rank $2$ Lie algebra (i.e. dimension of its Cartan subalgebra is equal to $2$). There are $2$ simple roots $\alpha,\,\beta$ and $6$ positive roots: $\alpha,\beta,\alpha+\beta,2\alpha+\beta,3\alpha+\beta,3\alpha+2\beta$ and as many negative ones (given by adding minus to the positive); so the real dimension of this algebra is $14$; it is spanned by $2$ Cartan vectors, $H_\alpha,H_\beta$, $6$ positive and $6$ negative root vectors. The commutation relations of the corresponding positive root vectors are determined by the formulas (see \cite{Gro}):
\beq{commrelationsTH}
\begin{aligned}
{}&[E_\alpha,E_\beta]=E_{\alpha+\beta}, & &[E_\alpha,E_{\alpha+\beta}]=2E_{2\alpha+\beta}, \\
&[E_\alpha,E_{2\alpha+\beta}]=3E_{3\alpha+\beta}, & &[E_\beta,E_{3\alpha+\beta}]=-E_{3\alpha+2\beta}, \\
&[E_{\alpha+\beta},E_{2\alpha+\beta}]=3E_{3\alpha+2\beta}.
\end{aligned}
\eq
The other relations follow from the definitions, see section \ref{sec:roots}. These root vectors are represented by the following matrices (see \cite{Gro}):
\begin{equation*}
\begin{aligned}
E_\alpha&=
\begin{pmatrix}
 0 & -\sqrt{2} & 0 & 0 & 0 & 0 & 0 \\
 0 & 0 & 0 & 0 & 0 & 0 & 0 \\
 0 & 0 & 0 & 0 & 0 & 0 & 1 \\
 0 & 0 & 0 & 0 & 0 & -1 & 0 \\
 \sqrt{2} & 0 & 0 & 0 & 0 & 0 & 0 \\
 0 & 0 & 0 & 0 & 0 & 0 & 0 \\
 0 & 0 & 0 & 0 & 0 & 0 & 0
\end{pmatrix}, &
E_{-\alpha}&=
\begin{pmatrix}
 0 & 0 & 0 & 0 & -\sqrt{2} & 0 & 0 \\
 \sqrt{2} & 0 & 0 & 0 & 0 & 0 & 0 \\
 0 & 0 & 0 & 0 & 0 & 0 & 0 \\
 0 & 0 & 0 & 0 & 0 & 0 & 0 \\
 0 & 0 & 0 & 0 & 0 & 0 & 0 \\
 0 & 0 & 0 & 1 & 0 & 0 & 0 \\
 0 & 0 & -1 & 0 & 0 & 0 & 0
\end{pmatrix}, \\
E_\beta&= \begin{pmatrix}
 0 & 0 & 0 & 0 & 0 & 0 & 0 \\
 0 & 0 & 1 & 0 & 0 & 0 & 0 \\
 0 & 0 & 0 & 0 & 0 & 0 & 0 \\
 0 & 0 & 0 & 0 & 0 & 0 & 0 \\
 0 & 0 & 0 & 0 & 0 & 0 & 0 \\
 0 & 0 & 0 & 0 & -1 & 0 & 0 \\
 0 & 0 & 0 & 0 & 0 & 0 & 0
\end{pmatrix}, &
E_{-\beta}&=
\begin{pmatrix}
 0 & 0 & 0 & 0 & 0 & 0 & 0 \\
 0 & 0 & 0 & 0 & 0 & 0 & 0 \\
 0 & 1 & 0 & 0 & 0 & 0 & 0 \\
 0 & 0 & 0 & 0 & 0 & 0 & 0 \\
 0 & 0 & 0 & 0 & 0 & -1 & 0 \\
 0 & 0 & 0 & 0 & 0 & 0 & 0 \\
 0 & 0 & 0 & 0 & 0 & 0 & 0
\end{pmatrix},\\
E_{\alpha+\beta}&=
\begin{pmatrix}
 0 & 0 & -\sqrt{2} & 0 & 0 & 0 & 0 \\
 0 & 0 & 0 & 0 & 0 & 0 & -1 \\
 0 & 0 & 0 & 0 & 0 & 0 & 0 \\
 0 & 0 & 0 & 0 & 1 & 0 & 0 \\
 0 & 0 & 0 & 0 & 0 & 0 & 0 \\
 \sqrt{2} & 0 & 0 & 0 & 0 & 0 & 0 \\
 0 & 0 & 0 & 0 & 0 & 0 & 0
\end{pmatrix}, &
E_{-(\alpha+\beta)}&=
\begin{pmatrix}
 0 & 0 & 0 & 0 & 0 & -\sqrt{2} & 0 \\
 0 & 0 & 0 & 0 & 0 & 0 & 0 \\
 \sqrt{2} & 0 & 0 & 0 & 0 & 0 & 0 \\
 0 & 0 & 0 & 0 & 0 & 0 & 0 \\
 0 & 0 & 0 & -1 & 0 & 0 & 0 \\
 0 & 0 & 0 & 0 & 0 & 0 & 0 \\
 0 & 1 & 0 & 0 & 0 & 0 & 0
\end{pmatrix},
\end{aligned}
\end{equation*}
\begin{equation*}
\begin{aligned}
E_{2\alpha+\beta}&=
\begin{pmatrix}
 0 & 0 & 0 & 0 & 0 & 0 & -\sqrt{2} \\
 0 & 0 & 0 & 0 & 0 & 0 & 0 \\
 0 & 0 & 0 & 0 & 0 & 0 & 0 \\
 \sqrt{2} & 0 & 0 & 0 & 0 & 0 & 0 \\
 0 & 0 & 1 & 0 & 0 & 0 & 0 \\
 0 & -1 & 0 & 0 & 0 & 0 & 0 \\
 0 & 0 & 0 & 0 & 0 & 0 & 0
\end{pmatrix}, &
E_{-(2\alpha+\beta)}&=
\begin{pmatrix}
 0 & 0 & 0 & -\sqrt{2} & 0 & 0 & 0 \\
 0 & 0 & 0 & 0 & 0 & 1 & 0 \\
 0 & 0 & 0 & 0 & -1 & 0 & 0 \\
 0 & 0 & 0 & 0 & 0 & 0 & 0 \\
 0 & 0 & 0 & 0 & 0 & 0 & 0 \\
 0 & 0 & 0 & 0 & 0 & 0 & 0 \\
 \sqrt{2} & 0 & 0 & 0 & 0 & 0 & 0
\end{pmatrix},\\
E_{3\alpha+\beta}&=
\begin{pmatrix}
 0 & 0 & 0 & 0 & 0 & 0 & 0 \\
 0 & 0 & 0 & 0 & 0 & 0 & 0 \\
 0 & 0 & 0 & 0 & 0 & 0 & 0 \\
 0 & 1 & 0 & 0 & 0 & 0 & 0 \\
 0 & 0 & 0 & 0 & 0 & 0 & -1 \\
 0 & 0 & 0 & 0 & 0 & 0 & 0 \\
 0 & 0 & 0 & 0 & 0 & 0 & 0
\end{pmatrix}, &
E_{-(3\alpha+\beta)}&=
\begin{pmatrix}
 0 & 0 & 0 & 0 & 0 & 0 & 0 \\
 0 & 0 & 0 & 1 & 0 & 0 & 0 \\
 0 & 0 & 0 & 0 & 0 & 0 & 0 \\
 0 & 0 & 0 & 0 & 0 & 0 & 0 \\
 0 & 0 & 0 & 0 & 0 & 0 & 0 \\
 0 & 0 & 0 & 0 & 0 & 0 & 0 \\
 0 & 0 & 0 & 0 & -1 & 0 & 0
\end{pmatrix}, \\
E_{3\alpha+2\beta}&=
\begin{pmatrix}
 0 & 0 & 0 & 0 & 0 & 0 & 0 \\
 0 & 0 & 0 & 0 & 0 & 0 & 0 \\
 0 & 0 & 0 & 0 & 0 & 0 & 0 \\
 0 & 0 & 1 & 0 & 0 & 0 & 0 \\
 0 & 0 & 0 & 0 & 0 & 0 & 0 \\
 0 & 0 & 0 & 0 & 0 & 0 & -1 \\
 0 & 0 & 0 & 0 & 0 & 0 & 0
\end{pmatrix}, &
E_{-(3\alpha+2\beta)}&=
\begin{pmatrix}
 0 & 0 & 0 & 0 & 0 & 0 & 0 \\
 0 & 0 & 0 & 0 & 0 & 0 & 0 \\
 0 & 0 & 0 & 1 & 0 & 0 & 0 \\
 0 & 0 & 0 & 0 & 0 & 0 & 0 \\
 0 & 0 & 0 & 0 & 0 & 0 & 0 \\
 0 & 0 & 0 & 0 & 0 & 0 & 0 \\
 0 & 0 & 0 & 0 & 0 & -1 & 0
\end{pmatrix}.
\end{aligned}
\end{equation*}
As one can see, this embedding does not verify the condition, that positive (and negative) root vectors are represented by strictly-triangular matrices, although $E_\gamma$ and $E_{-\gamma}$ are represented by transposed matrices. In order to rule out this flaw, we conjugate the representation by the orthogonal matrix
\[
P=
\begin{pmatrix}
0 & 1 & 0 & 0 & 0 & 0 & 0 \\
 0 & 0 & 0 & 0 & 0 & 1 & 0 \\
 0 & 0 & 0 & 0 & 0 & 0 & 1 \\
 1 & 0 & 0 & 0 & 0 & 0 & 0 \\
 0 & 0 & 0 & 1 & 0 & 0 & 0 \\
 0 & 0 & 1 & 0 & 0 & 0 & 0 \\
 0 & 0 & 0 & 0 & 1 & 0 & 0
 \end{pmatrix}
\]
and put
\begin{equation}
\label{rvect}
E^{new}_\gamma = (PE_{-\gamma}P^{-1})^T, \ \ \ E^{new}_{-\delta} = (PE_\delta P^{-1})^T.
\end{equation}
Since $P$ is orthogonal this transformation does not affect the crucial property ($E_{-\gamma}=E_\gamma^T$). In this way we obtain the following representation for the root vectors:
\begin{equation*}
\begin{aligned}
E^{new}_\alpha &=
\begin{pmatrix}
 0 & 0 & 0 & 0 & 0 & 0 & 0 \\
 0 & 0 & 0 & 0 & 0 & 0 & 0 \\
 0 & 0 & 0 & 0 & 0 & 0 & 0 \\
 -\sqrt{2} & 0 & 0 & 0 & 0 & 0 & 0 \\
 0 & -1 & 0 & 0 & 0 & 0 & 0 \\
 0 & 0 & 1 & 0 & 0 & 0 & 0 \\
 0 & 0 & 0 & \sqrt{2} & 0 & 0 & 0
\end{pmatrix} &
E^{new}_{-\alpha}&=
\begin{pmatrix}
 0 & 0 & 0 & -\sqrt{2} & 0 & 0 & 0 \\
 0 & 0 & 0 & 0 & -1 & 0 & 0 \\
 0 & 0 & 0 & 0 & 0 & 1 & 0 \\
 0 & 0 & 0 & 0 & 0 & 0 & \sqrt{2} \\
 0 & 0 & 0 & 0 & 0 & 0 & 0 \\
 0 & 0 & 0 & 0 & 0 & 0 & 0 \\
 0 & 0 & 0 & 0 & 0 & 0 & 0
\end{pmatrix}
\end{aligned}
\end{equation*}
\begin{equation*}
\begin{aligned}
E^{new}_\beta&=
\begin{pmatrix}
 0 & 0 & 0 & 0 & 0 & 1 & 0 \\
 0 & 0 & 0 & 0 & 0 & 0 & -1 \\
 0 & 0 & 0 & 0 & 0 & 0 & 0 \\
 0 & 0 & 0 & 0 & 0 & 0 & 0 \\
 0 & 0 & 0 & 0 & 0 & 0 & 0 \\
 0 & 0 & 0 & 0 & 0 & 0 & 0 \\
 0 & 0 & 0 & 0 & 0 & 0 & 0
\end{pmatrix} &
E^{new}_{-\beta}&=
\begin{pmatrix}
 0 & 0 & 0 & 0 & 0 & 0 & 0 \\
 0 & 0 & 0 & 0 & 0 & 0 & 0 \\
 0 & 0 & 0 & 0 & 0 & 0 & 0 \\
 0 & 0 & 0 & 0 & 0 & 0 & 0 \\
 0 & 0 & 0 & 0 & 0 & 0 & 0 \\
 1 & 0 & 0 & 0 & 0 & 0 & 0 \\
 0 & -1 & 0 & 0 & 0 & 0 & 0
\end{pmatrix}\\
E^{new}_{\alpha+\beta}&=
\begin{pmatrix}
 0 & 0 & -1 & 0 & 0 & 0 & 0 \\
 0 & 0 & 0 & \sqrt{2} & 0 & 0 & 0 \\
 0 & 0 & 0 & 0 & 0 & 0 & 0 \\
 0 & 0 & 0 & 0 & 0 & -\sqrt{2} & 0 \\
 0 & 0 & 0 & 0 & 0 & 0 & 1 \\
 0 & 0 & 0 & 0 & 0 & 0 & 0 \\
 0 & 0 & 0 & 0 & 0 & 0 & 0
\end{pmatrix} &
E^{new}_{-(\alpha+\beta)}&=
\begin{pmatrix}
 0 & 0 & 0 & 0 & 0 & 0 & 0 \\
 0 & 0 & 0 & 0 & 0 & 0 & 0 \\
 -1 & 0 & 0 & 0 & 0 & 0 & 0 \\
 0 & \sqrt{2} & 0 & 0 & 0 & 0 & 0 \\
 0 & 0 & 0 & 0 & 0 & 0 & 0 \\
 0 & 0 & 0 & -\sqrt{2} & 0 & 0 & 0 \\
 0 & 0 & 0 & 0 & 1 & 0 & 0
\end{pmatrix} \\
E^{new}_{2\alpha+\beta}&=
\begin{pmatrix}
 0 & 0 & 0 & 0 & 0 & 0 & 0 \\
 -1 & 0 & 0 & 0 & 0 & 0 & 0 \\
 0 & 0 & 0 & 0 & 0 & 0 & 0 \\
 0 & 0 & -\sqrt{2} & 0 & 0 & 0 & 0 \\
 0 & 0 & 0 & \sqrt{2} & 0 & 0 & 0 \\
 0 & 0 & 0 & 0 & 0 & 0 & 0 \\
 0 & 0 & 0 & 0 & 0 & 1 & 0
\end{pmatrix} &
E^{new}_{-(2\alpha+\beta)}&=
\begin{pmatrix}
 0 & -1 & 0 & 0 & 0 & 0 & 0 \\
 0 & 0 & 0 & 0 & 0 & 0 & 0 \\
 0 & 0 & 0 & -\sqrt{2} & 0 & 0 & 0 \\
 0 & 0 & 0 & 0 & \sqrt{2} & 0 & 0 \\
 0 & 0 & 0 & 0 & 0 & 0 & 0 \\
 0 & 0 & 0 & 0 & 0 & 0 & 1 \\
 0 & 0 & 0 & 0 & 0 & 0 & 0
\end{pmatrix}\\
E^{new}_{3\alpha+\beta}&=
\begin{pmatrix}
 0 & 0 & 0 & 0 & 0 & 0 & 0 \\
 0 & 0 & 0 & 0 & 0 & 0 & 0 \\
 0 & 0 & 0 & 0 & 0 & 0 & 0 \\
 0 & 0 & 0 & 0 & 0 & 0 & 0 \\
 -1 & 0 & 0 & 0 & 0 & 0 & 0 \\
 0 & 0 & 0 & 0 & 0 & 0 & 0 \\
 0 & 0 & 1 & 0 & 0 & 0 & 0
\end{pmatrix} &
E^{new}_{-(3\alpha+\beta)}&=
\begin{pmatrix}
 0 & 0 & 0 & 0 & -1 & 0 & 0 \\
 0 & 0 & 0 & 0 & 0 & 0 & 0 \\
 0 & 0 & 0 & 0 & 0 & 0 & 1 \\
 0 & 0 & 0 & 0 & 0 & 0 & 0 \\
 0 & 0 & 0 & 0 & 0 & 0 & 0 \\
 0 & 0 & 0 & 0 & 0 & 0 & 0 \\
 0 & 0 & 0 & 0 & 0 & 0 & 0
\end{pmatrix}\\
E^{new}_{3\alpha+2\beta}&=
\begin{pmatrix}
 0 & 0 & 0 & 0 & 0 & 0 & 0 \\
 0 & 0 & 1 & 0 & 0 & 0 & 0 \\
 0 & 0 & 0 & 0 & 0 & 0 & 0 \\
 0 & 0 & 0 & 0 & 0 & 0 & 0 \\
 0 & 0 & 0 & 0 & 0 & -1 & 0 \\
 0 & 0 & 0 & 0 & 0 & 0 & 0 \\
 0 & 0 & 0 & 0 & 0 & 0 & 0
\end{pmatrix} &
E^{new}_{-(3\alpha+2\beta)}&=
\begin{pmatrix}
 0 & 0 & 0 & 0 & 0 & 0 & 0 \\
 0 & 0 & 0 & 0 & 0 & 0 & 0 \\
 0 & 1 & 0 & 0 & 0 & 0 & 0 \\
 0 & 0 & 0 & 0 & 0 & 0 & 0 \\
 0 & 0 & 0 & 0 & 0 & 0 & 0 \\
 0 & 0 & 0 & 0 & -1 & 0 & 0 \\
 0 & 0 & 0 & 0 & 0 & 0 & 0
\end{pmatrix}
\end{aligned}
\end{equation*}
These matrices also verify the commutator relations of $G_2$ type, although a little modified (due to the renaming, see \eqref{rvect}):
\begin{equation*}
\begin{aligned}
{}&[E^{new}_\alpha,E^{new}_\beta]=E^{new}_{\alpha+\beta}, & &[E^{new}_\alpha,E^{new}_{\alpha+\beta}]=2E^{new}_{2\alpha+\beta},\\
{}&[E^{new}_\alpha,E^{new}_{2\alpha+\beta}]=-3E^{new}_{3\alpha+\beta}, & &[E^{new}_\beta,E^{new}_{3\alpha+\beta}]=-E^{new}_{3\alpha+2\beta}, \\
{}&[E^{new}_{\alpha+\beta},E^{new}_{2\alpha+\beta}]=-3E^{new}_{3\alpha+2\beta}.
\end{aligned}
\end{equation*}
As one sees all the matrices are either strictly upper or strictly lower-triangular here, so that the matrices, corresponding to the roots $+\gamma$ and $-\gamma$ (for all $\gamma$) are transposes of each other. Thus the conditions on the embedding that one needs to implement the usual construction of Toda system, hold. In effect, one can make all the matrices, corresponding to positive roots upper triangular: to this end one should rename basis roots, but we do not need this in our work.

The Cartan algebra in this case (we obtain it by conjugating by $P$ the algebra, considered in \cite{Gro}) consists of the matrices
\[
\Lambda=\left(
\begin{array}{ccccccc}
 \lambda _1 & 0 & 0 & 0 & 0 & 0 & 0 \\
 0 & -\lambda _2 & 0 & 0 & 0 & 0 & 0 \\
 0 & 0 & -\lambda _3 & 0 & 0 & 0 & 0 \\
 0 & 0 & 0 & 0 & 0 & 0 & 0 \\
 0 & 0 & 0 & 0 & \lambda _3 & 0 & 0 \\
 0 & 0 & 0 & 0 & 0 & \lambda _2 & 0 \\
 0 & 0 & 0 & 0 & 0 & 0 &- \lambda _1
\end{array}
\right),
\]
where $\lambda_1+\lambda_2+\lambda_3=0$ and one can write down explicit matrix representatives of the Weyl group elements in this representation (i.e. the elements in the maximal compact subgroup of $G_2\subset SL(7,\mathbb R)$, such that conjugation of the Cartan matrices by these elements gives the action of Weyl group):
\begin{equation*}
\begin{aligned}
\tilde\omega_1&=
\begin{pmatrix}
 0 & 0 & 0 & 0 & 0 & 0 & 1 \\
 0 & 0 & 0 & 0 & 1 & 0 & 0 \\
 0 & 0 & 0 & 0 & 0 & 1 & 0 \\
 0 & 0 & 0 & -1 & 0 & 0 & 0 \\
 0 & 1 & 0 & 0 & 0 & 0 & 0 \\
 0 & 0 & 1 & 0 & 0 & 0 & 0 \\
 1 & 0 & 0 & 0 & 0 & 0 & 0
\end{pmatrix}, &
\tilde\omega_2&=
\begin{pmatrix}
 0 & 0 & 0 & 0 & 0 & 1 & 0 \\
 0 & 0 & 0 & 0 & 0 & 0 & -1 \\
 0 & 0 & 1 & 0 & 0 & 0 & 0 \\
 0 & 0 & 0 & 1 & 0 & 0 & 0 \\
 0 & 0 & 0 & 0 & 1 & 0 & 0 \\
 -1 & 0 & 0 & 0 & 0 & 0 & 0 \\
 0 & 1 & 0 & 0 & 0 & 0 & 0
\end{pmatrix}, \\
\tilde\omega_3&=
\begin{pmatrix}
 0 & 1 & 0 & 0 & 0 & 0 & 0 \\
 0 & 0 & 0 & 0 & 1 & 0 & 0 \\
 -1 & 0 & 0 & 0 & 0 & 0 & 0 \\
 0 & 0 & 0 & -1 & 0 & 0 & 0 \\
 0 & 0 & 0 & 0 & 0 & 0 & -1 \\
 0 & 0 & 1 & 0 & 0 & 0 & 0 \\
 0 & 0 & 0 & 0 & 0 & 1 & 0
\end{pmatrix}, &
\tilde\omega_4&=
\begin{pmatrix}
 0 & 0 & 1 & 0 & 0 & 0 & 0 \\
 0 & 0 & 0 & 0 & 0 & -1 & 0 \\
 -1 & 0 & 0 & 0 & 0 & 0 & 0 \\
 0 & 0 & 0 & -1 & 0 & 0 & 0 \\
 0 & 0 & 0 & 0 & 0 & 0 & -1 \\
 0 & -1 & 0 & 0 & 0 & 0 & 0 \\
 0 & 0 & 0 & 0 & 1 & 0 & 0
\end{pmatrix}, \\
\tilde\omega_5&=
\begin{pmatrix}
 0 & 0 & 0 & 0 & 1 & 0 & 0 \\
 0 & 0 & 0 & 0 & 0 & 0 & -1 \\
 0 & -1 & 0 & 0 & 0 & 0 & 0 \\
 0 & 0 & 0 & 1 & 0 & 0 & 0 \\
 0 & 0 & 0 & 0 & 0 & -1 & 0 \\
 -1 & 0 & 0 & 0 & 0 & 0 & 0 \\
 0 & 0 & 1 & 0 & 0 & 0 & 0
\end{pmatrix}, &
\tilde\omega_6&=
\begin{pmatrix}
 -1 & 0 & 0 & 0 & 0 & 0 & 0 \\
 0 & 0 & -1 & 0 & 0 & 0 & 0 \\
 0 & -1 & 0 & 0 & 0 & 0 & 0 \\
 0 & 0 & 0 & 1 & 0 & 0 & 0 \\
 0 & 0 & 0 & 0 & 0 & -1 & 0 \\
 0 & 0 & 0 & 0 & -1 & 0 & 0 \\
 0 & 0 & 0 & 0 & 0 & 0 & -1
\end{pmatrix}, \\
\tilde\omega_7&=
\begin{pmatrix}
 0 & 0 & 0 & 0 & 0 & 0 & -1 \\
 0 & 0 & 0 & 0 & 0 & -1 & 0 \\
 0 & 0 & 0 & 0 & -1 & 0 & 0 \\
 0 & 0 & 0 & -1 & 0 & 0 & 0 \\
 0 & 0 & -1 & 0 & 0 & 0 & 0 \\
 0 & -1 & 0 & 0 & 0 & 0 & 0 \\
 -1 & 0 & 0 & 0 & 0 & 0 & 0
\end{pmatrix}, &
\tilde\omega_8&=
\begin{pmatrix}
 0 & -1 & 0 & 0 & 0 & 0 & 0 \\
 1 & 0 & 0 & 0 & 0 & 0 & 0 \\
 0 & 0 & 0 & 0 & -1 & 0 & 0 \\
 0 & 0 & 0 & -1 & 0 & 0 & 0 \\
 0 & 0 & -1 & 0 & 0 & 0 & 0 \\
 0 & 0 & 0 & 0 & 0 & 0 & 1 \\
 0 & 0 & 0 & 0 & 0 & -1 & 0
\end{pmatrix},
\end{aligned}
\end{equation*}
\begin{equation*}
\begin{aligned}
\tilde\omega_9&=
\begin{pmatrix}
 0 & 0 & 0 & 0 & 0 & -1 & 0 \\
 0 & 0 & -1 & 0 & 0 & 0 & 0 \\
 0 & 0 & 0 & 0 & 0 & 0 & 1 \\
 0 & 0 & 0 & 1 & 0 & 0 & 0 \\
 1 & 0 & 0 & 0 & 0 & 0 & 0 \\
 0 & 0 & 0 & 0 & -1 & 0 & 0 \\
 0 & -1 & 0 & 0 & 0 & 0 & 0
\end{pmatrix}, &
\tilde\omega_{10}&=
\begin{pmatrix}
 0 & 0 & 0 & 0 & -1 & 0 & 0 \\
 0 & 1 & 0 & 0 & 0 & 0 & 0 \\
 0 & 0 & 0 & 0 & 0 & 0 & 1 \\
 0 & 0 & 0 & 1 & 0 & 0 & 0 \\
 1 & 0 & 0 & 0 & 0 & 0 & 0 \\
 0 & 0 & 0 & 0 & 0 & 1 & 0 \\
 0 & 0 & -1 & 0 & 0 & 0 & 0
\end{pmatrix}, \\
\tilde\omega_{11}&=
\begin{pmatrix}
 0 & 0 & -1 & 0 & 0 & 0 & 0 \\
 1 & 0 & 0 & 0 & 0 & 0 & 0 \\
 0 & 0 & 0 & 0 & 0 & 1 & 0 \\
 0 & 0 & 0 & -1 & 0 & 0 & 0 \\
 0 & 1 & 0 & 0 & 0 & 0 & 0 \\
 0 & 0 & 0 & 0 & 0 & 0 & 1 \\
 0 & 0 & 0 & 0 & -1 & 0 & 0
\end{pmatrix}, &
\tilde\omega_0&=
\begin{pmatrix}
 1 & 0 & 0 & 0 & 0 & 0 & 0 \\
 0 & 1 & 0 & 0 & 0 & 0 & 0 \\
 0 & 0 & 1 & 0 & 0 & 0 & 0 \\
 0 & 0 & 0 & 1 & 0 & 0 & 0 \\
 0 & 0 & 0 & 0 & 1 & 0 & 0 \\
 0 & 0 & 0 & 0 & 0 & 1 & 0 \\
 0 & 0 & 0 & 0 & 0 & 0 & 1
\end{pmatrix}.
\end{aligned}
\end{equation*}
Observe, that since we are only interested in the permutations of $\pm\lambda_i$, induced by the conjugations by these matrices, the signs of the entries of $\tilde\omega_i$ here are irrelevant, so we could replace all $\pm1$ by asterisks in this table.

Now the remaining part of the reasoning is quite similar to the case of $Sp(4,\mathbb R)$: we consider the Morse function:
\[
F_{G_2}(L)=Tr(LN),
\]
where we can take (see \cite{BBR})
\[
N=\begin{pmatrix}
-11 & 0 & 0 & 0 & 0 & 0 & 0 \\
 0 & -8 & 0 & 0 & 0 & 0 & 0 \\
 0 & 0 & -3 & 0 & 0 & 0 & 0 \\
 0 & 0 & 0 & 0 & 0 & 0 & 0 \\
 0 & 0 & 0 & 0 & 3 & 0 & 0 \\
 0 & 0 & 0 & 0 & 0 & 8 & 0 \\
 0 & 0 & 0 & 0 & 0 & 0 & 11
\end{pmatrix}
\]
and $L$ is a symmetric matrix from $\mathfrak g_2$, which one can regard as the Lie subalgebra of $\mathfrak{sl}(7,\mathbb R)$, generated by the root matrices $E_\gamma^{new}$, listed above. Further, the antisymmetric part of this embedded algebra $\mathfrak g_2$, which corresponds to the Lie algebra of the maximal compact subgroup of $G_2$, is spanned by the matrices
\[
\Theta=\begin{pmatrix}
 0 & \theta _6 & -\theta _5 & -\sqrt{2} \theta _4 & \theta _2 &
   \theta _1 & 0 \\
 -\theta _6 & 0 & -\theta _3 & \sqrt{2} \theta _5 & -\theta _4 & 0
   & -\theta _1 \\
 \theta _5 & \theta _3 & 0 & \sqrt{2} \theta _6 & 0 & \theta _4 &
   -\theta _2 \\
 \sqrt{2} \theta _4 & -\sqrt{2} \theta _5 & -\sqrt{2} \theta _6 &
   0 & -\sqrt{2} \theta _6 & -\sqrt{2} \theta _5 & \sqrt{2} \theta
   _4 \\
 -\theta _2 & \theta _4 & 0 & \sqrt{2} \theta _6 & 0 & \theta _3 &
   \theta _5 \\
 -\theta _1 & 0 & -\theta _4 & \sqrt{2} \theta _5 & -\theta _3 & 0
   & -\theta _6 \\
 0 & \theta _1 & \theta _2 & -\sqrt{2} \theta _4 & -\theta _5 &
   \theta _6 & 0
\end{pmatrix},
\]
where $\theta_i$ are (real) coordinates on this Lie algebra. We can use their left translations by $g\in G_2$ the (local) coordinates on $G_2$ in a neighbourhood of $g$. Then the quadratic part of $F_{G_2}$ at the singular points, represented by the Weyl group representatives $\tilde\omega_i$, listed above, can be explicitly computed. For example:
\begin{equation*}
\begin{aligned}
\frac{1}{2}d^2_{\tilde{\omega}_{0}}F_{G_2}&=19 \theta _1^2 \left(\lambda _1-\lambda _2\right)+14 \theta _2^2 \left(\lambda _1-\lambda _3\right)\\
                                                                    &\quad+5 \theta _3^2 \left(\lambda _3-\lambda _2\right)+11 \theta _4^2 \left(2 \lambda _1-\lambda _2-\lambda _3\right)\\
                                                                    &+8 \theta _5^2 \left(\lambda _1-2\lambda _2+\lambda _3\right)
+3 \theta _6^2 \left(\lambda _1+\lambda _2-2 \lambda _3\right).
\end{aligned}
\end{equation*}
Now we fix the order of eigenvalues by prescribing, that $\lambda_1>\lambda_2>\lambda_3$. As before, one can compute the indices of the singular points, find the positive/negative with respect to the quadratic form $d^2F_{G_2}$ directions in the tangent space of the maximal compact subgroup of $G_2$ at these points, and list the minor surfaces, that contain these points. We put these data in a table:
\begin{equation*}
\footnotesize{
\begin{tabular}{|c|c|c|c|c|}
\hline\cline{1-0}
$$ & $$ & $$ & $$ & $$\\
$i$ & $\tilde{\omega_i}$ & $\theta _1,\theta _2,\theta _3,\theta _4,\theta _5,\theta _6$ & Index & Minors, $o_{ij}$\\

$$ & $$ & $$ & $$ & $$\\
\hline\cline{1-0}
$$ & $$ & $$ & $$ & $$\\
$1$ & $\tilde \omega_{0}$ & $+,+,-,+,-,+$ & $2$ & $o_{12},o_{13},o_{14},o_{15},o_{16},o_{17}$\\

$$ & $$ & $$ & $$ & $$\\
\hline\cline{1-0}
$$ & $$ & $$ & $$ & $$\\
$2$ & $\tilde \omega_{1}$ & $-,-,-,-,-,+$ & $5$ & $o_{11},o_{12},o_{13},o_{14},o_{15},o_{16}$\\

$$ & $$ & $$ & $$ & $$\\
\hline\cline{1-0}
$$ & $$ & $$ & $$ & $$\\
$3$ & $\tilde \omega_{2}$ & $-,+,-,+,-,+$ & $3$ & $o_{11},o_{12},o_{13},o_{14},o_{15},o_{17}$\\

$$ & $$ & $$ & $$ & $$\\
\hline\cline{1-0}
$$ & $$ & $$ & $$ & $$\\
$4$ & $\tilde \omega_{3}$ & $-,+,-,-,-,+$ & $4$ & $o_{11},o_{13},o_{14},o_{15},o_{16},o_{17}$\\

$$ & $$ & $$ & $$ & $$\\
\hline\cline{1-0}
$$ & $$ & $$ & $$ & $$\\
$5$ & $\tilde \omega_{4}$ & $+,+,-,+,+,+$ & $1$ & $o_{11},o_{12},o_{14},o_{15},o_{16},o_{17}$\\

$$ & $$ & $$ & $$ & $$\\
\hline\cline{1-0}
$$ & $$ & $$ & $$ & $$\\
$6$ & $\tilde \omega_{5}$ & $-,-,-,-,-,-$ & $6$ & $o_{11},o_{12},o_{13},o_{14},o_{16},o_{17}$\\

$$ & $$ & $$ & $$ & $$\\
\hline\cline{1-0}
$$ & $$ & $$ & $$ & $$\\
$7$ & $\tilde \omega_{6}$ & $+,+,+,+,+,-$ & $1$ & $o_{12},o_{13},o_{14},o_{15},o_{16},o_{17}$\\

$$ & $$ & $$ & $$ & $$\\
\hline\cline{1-0}
$$ & $$ & $$ & $$ & $$\\
$8$ & $\tilde \omega_{7}$ & $-,-,+,-,+,-$ & $4$ & $o_{11},o_{12},o_{13},o_{14},o_{15},o_{16}$\\

$$ & $$ & $$ & $$ & $$\\
\hline\cline{1-0}
$$ & $$ & $$ & $$ & $$\\
$9$ & $\tilde \omega_{8}$ & $+,-,+,-,+,-$ & $3$ & $o_{11},o_{13},o_{14},o_{15},o_{16},o_{17}$\\

$$ & $$ & $$ & $$ & $$\\
\hline\cline{1-0}
$$ & $$ & $$ & $$ & $$\\
$10$ & $\tilde \omega_{9}$ & $+,-,+,+,+,-$ & $2$ & $o_{11},o_{12},o_{13},o_{14},o_{15},o_{17}$\\

$$ & $$ & $$ & $$ & $$\\
\hline\cline{1-0}
$$ & $$ & $$ & $$ & $$\\
$11$ & $\tilde \omega_{10}$ & $-,-,+,-,-,-$ & $5$ & $o_{11},o_{12},o_{13},o_{14},o_{16},o_{17}$\\

$$ & $$ & $$ & $$ & $$\\
\hline\cline{1-0}
$$ & $$ & $$ & $$ & $$\\
$12$ & $\tilde \omega_{11}$ & $+,+,+,+,+,+$ & $0$ & $o_{11},o_{12},o_{14},o_{15},o_{16},o_{17}$\\

$$ & $$ & $$ & $$ & $$\\
\hline
\end{tabular}
\
}
\end{equation*}
Analyzing the invariant minor subvarieties we get the picture of all possible trajectories, connecting two singular points (see figure 1, at which we have omitted the ``maximal'' trajectory, going from the very bottom to the very top of the diagram). If we remove the lines which correspond to the higher-dimensional submanifolds, spanned by the trajectories, we get Fig.2. This picture coincides with the Hasse diagram of the Bruhat order of $G_{2}$ up to a permutation of elements: this is due to the fact, that the order of eigenvalues we chose is not the canonical one, which should be $\lambda_1<-\lambda_2<-\lambda_3<0<\lambda_3<\lambda_2<-\lambda_1$. One can check, that if this is taken into consideration, we obtain the usual Bruhat diagram of $D_6=W_{G_2}$; moreover, indices of the points once again coincide with the lengths of Weyl elements.
\begin{figure}[t!]
\begin{center}
\includegraphics[width=250pt,height=270pt]{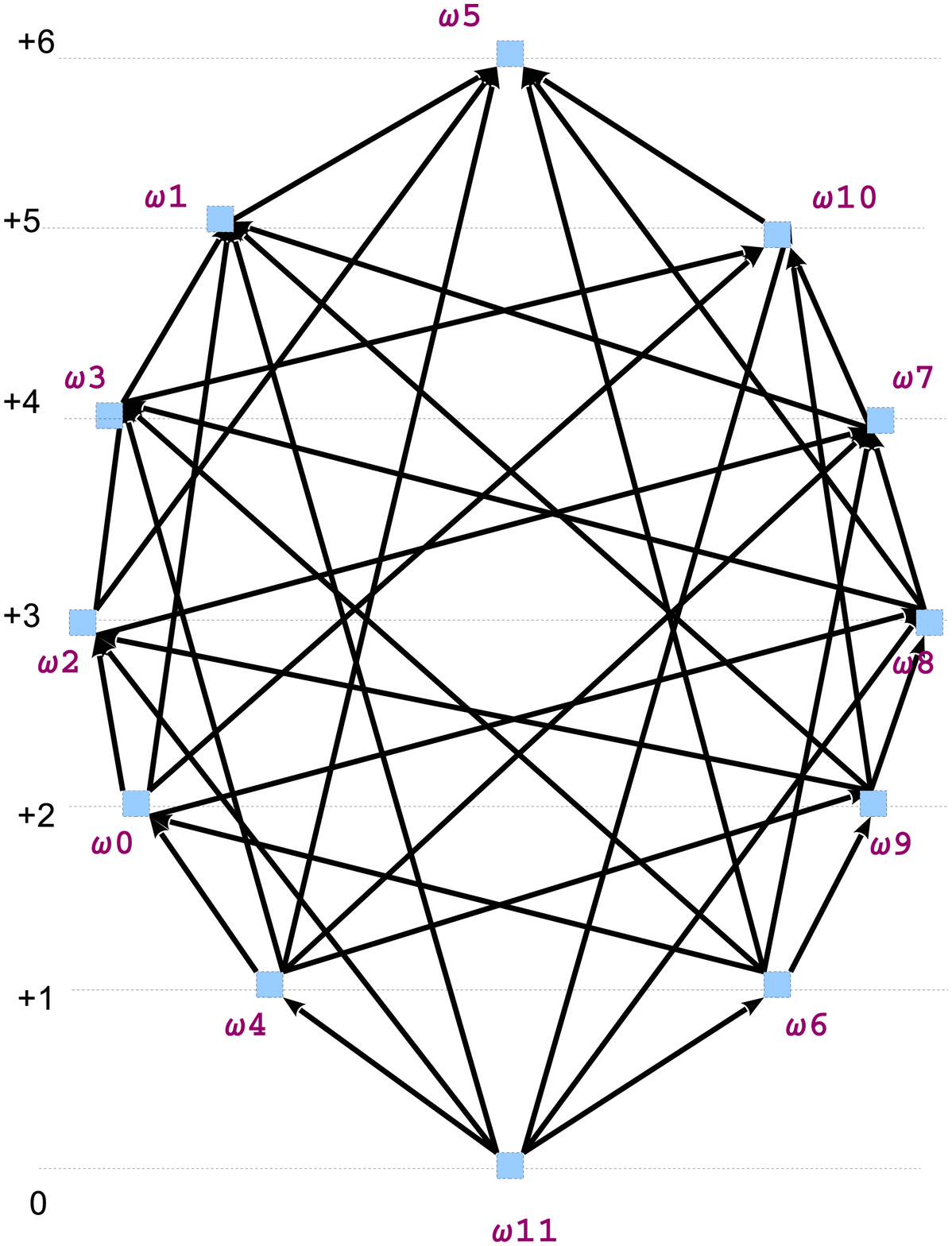}
\caption{All possible 1-d trajectories corresponding to the local coordinates $\theta_i$ and connecting two singular points}
\includegraphics[width=230pt,height=270pt]{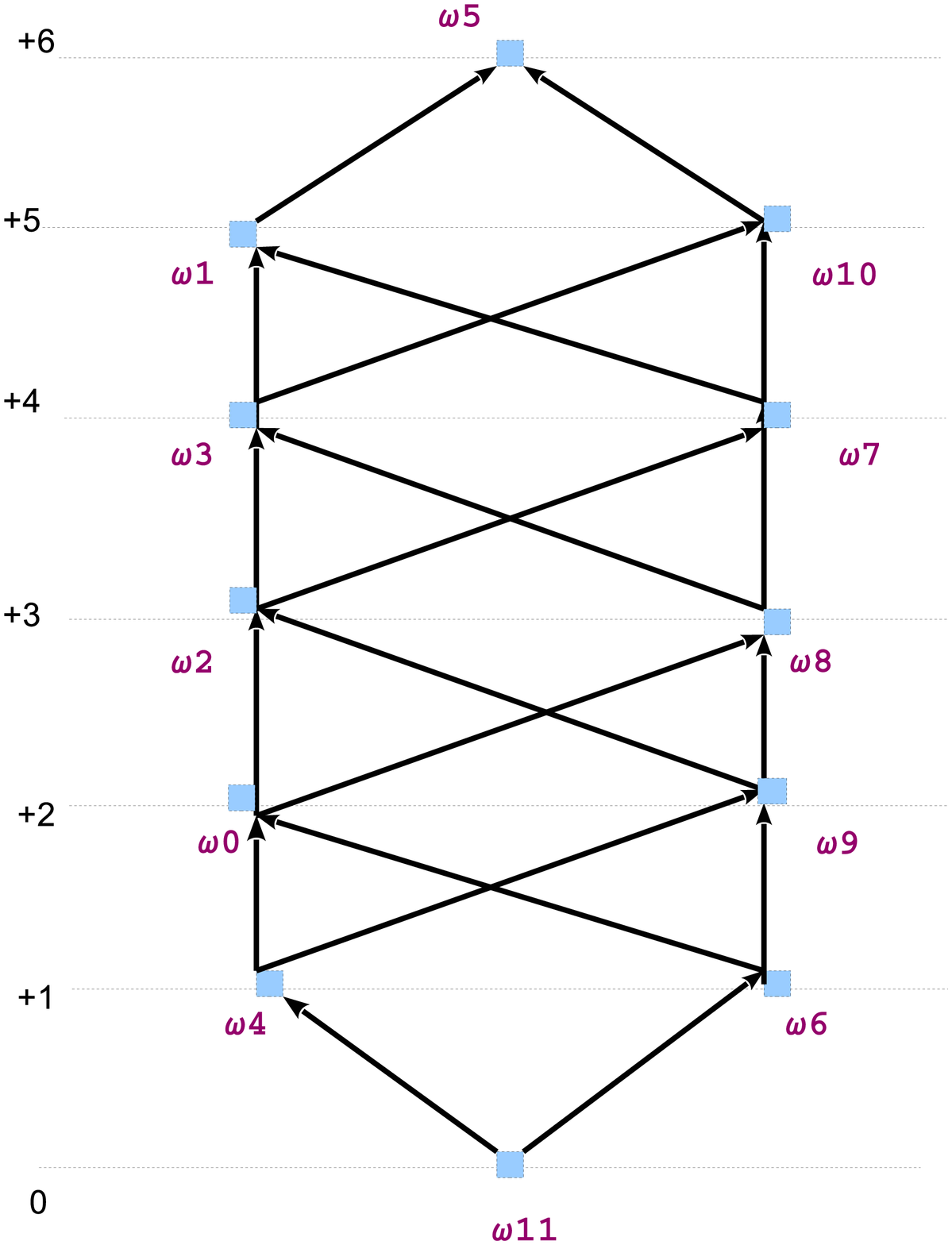}
\caption{1-d trajectories connecting two singular points, which are not covered by the trajectories of the higher dimensions.}
\end{center}
\end{figure}

\section{Conclusions}
This paper is the third in the series of our works (see \cite{CSS14}, \cite{CSS15}), devoted to the structure of the trajectories of generalized full symmetric Toda system. In our previous works we showed, that the phase portraits of this system on the $SL(n,\mathbb R)$ groups coincide with the Bruhat diagram of permutations, either in the usual sense if eigenvalues are distinct, or of permutations of multisets (when some eigenvalues coincide). As we have explained in the beginning, generalized Toda system can be defined for arbitrary semisimple Lie group. In this case Weyl group will play the role of the permutation group $S_n$. Thus the natural question is, whether similar result holds for in this case. More accurately, we come up with the following conjecture:
\begin{conj}
The phase picture of the Toda system associated with a semisimple Lie group when all eigenvalues are distinct coincides with the Bruhat order on the corresponding Weyl group. More accurately, we claim that the stationary points of this system correspond bijectively to the elements of the Weyl group, and that there is a trajectory between two such points if and only if the corresponding elements are comparable in Bruhat order. Moreover, the dimension of the space of such trajectories is equal to the number of points in the corresponding Bruhat interval. In particular, the Morse index of the point is equal to the length of the element in the Weyl group. Similar statement holds for the case of coinciding eigenvalues, but this time one should look at the Bruhat order on the corresponding ``partial flag space''.
\end{conj}

This statement looks very plausible. However, it is not so clear, if it is easy to prove it in a generic case. One of the possible strategy could be to use the inclusion of a semisimple group into $SL(n;\mathbb R)$ (for suitable $n$), which we have used here. Then the Toda system dynamics is induced by a restriction of the equation \eqref{LM} to the embedded group, so the stable points of this system can be found, as well as a certain collection of invariant subvarieties, induced by the intersections of the embedded subgroup with invariant varieties of the Toda flow on $SL(n,\mathbb R)$, see \cite{CS}, \cite{CS2}, \cite{CSS14}.

This, however is not enough to make up a final conclusion, since not all the necessary ingredients of our reasonings from \cite{CSS14} can be transferred from the big group to the embedded subgroup in this way. For instance, a direct computation shows, that the Morse function, that induces Toda system associated with $Sp(4;\mathbb R)$, does not coincide with the restriction of the Morse function of $SL(4;\mathbb R)$ (i.e the function that induces the full symmetric Toda system). This means, that one cannot use the properties of the Morse function on $SL(n,\mathbb R)$ (its stable points, hessians etc.) in the study of the system on a subgroup. On the other hand, there are more than one embedding $Sp(4,\mathbb R)\to SL(n,\mathbb R)$ that verify the conditions we listed above (i.e. sends the Cartan algebra into diagonal matrices and Borel subgroups into upper/lower triangular matrices), and we cannot say, if some of them verify additional properties or not.

It is even less clear, how these embeddings behave with respect to Bruhat cells. For instance, it is not clear, if such embedding would send Bruhat cells of the subgroup into the cells of $SL(n;\mathbb R)$; i.e. it looks quite plausible, that the embedding we need can be chosen so that Bruhat cells in the subgroup are equal to the intersections of this subgroup and Bruhat cells in $SL(n;\mathbb R)$. If this is so, then there would be a clear strategy of proving the conjecture, since we know (see \cite{CSS14}) that Bruhat cells in $SL(n,\mathbb R)$ are invariant varieties of the full symmetric Toda system. The plausibility of this statement is supported by a simple diagramm search, that shows, that the order, induced on the Weyl group of $Sp(4;\mathbb R)$ from its natural inclusion into $S_4$ coincides with the intrinsic Bruhat order on it (even though the lengths of the elements in these two groups do not coincide).

However, the question, whether the embedding that intertwines the Bruhat cells exists, although quite classical as it is formulated, seems to be not known to specialists. It seems to be quite difficult to prove this statement directly, since the very notion of Cartan algebra of $SL(n,\mathbb R)$ differs a lot from the image of Cartan algebra of the included subgroup, even when the inclusion sends it into diagonal matrices in $\mathfrak{sl}(n,\mathbb R)$. For example, in the case of $\mathfrak{sp}(4,\mathbb R)\subset\mathfrak{sl}(4,\mathbb R)$, the image of the Cartan subalgebra on the left consists of the matrices whose eigenvalues have multiplicity $2$, while from the point of view of the symplectic algebra, their coordinates are distinct. It also turns out that rather little is known about the combinatorics of Bruhat order on simple groups other than $SL(n,\mathbb R)$ and its relation with the Bruhat cells and dual Bruhat cells on the groups and flag spaces. All this makes the problem more difficult.

On the other hand, one may hope that solving this problem might shed some light on the structure of Bruhat order on Weyl groups: for example one would be able to use the Toda dynamics (which is rather well-studied) to describe the order in the cases, where the usual way to compute this order is too complicated. The structures of ``partial flag manifolds'', one obtains by letting some of eigenvalues coincide, seem to be even more intriguing.

We hope to solve these problems in future.

\paragraph{Acknowledgments}
The work of Yu.B. Chernyakov was supported by grant RFBR-15-02-04175. The work of G.I Sharygin was supported by grant RFBR-15-01-05990. The work of A.S. Sorin  was partially supported by the RFBR Grants No. 16-52-12012 -NNIO-a, No. 15-52-05022-Arm-a
and by the DFG Grant LE 838/12-2.

\end{document}